\documentclass[hidelinks,aps,pra,preprint,superscriptaddress,10pt]{revtex4-2} 

\usepackage{amsmath,amssymb}
\usepackage{mathrsfs}
\usepackage{epstopdf}
\usepackage{graphicx}
\usepackage{bm,color,bbm}
\usepackage{natbib}
\usepackage[hidelinks]{hyperref}
\usepackage{caption}
\usepackage{subcaption}
\usepackage{fancyhdr}
\usepackage{array}
\newcommand{\nn}{\nonumber}

\newcolumntype{P}[1]{>{\centering\arraybackslash}p{#1}}
\newcolumntype{M}[1]{>{\centering\arraybackslash}m{#1}}

\begin{document}
\fancyhead{} 
\fancyhead[LE,RO]{\ifnum\value{page}<2\relax\else\thepage\fi}

\title{Comments on: ``Introduction to the Absolute Brightness and Number Statistics in Spontaneous Parametric Down-Conversion" (2019 J. Opt. 21, 043501)}

\author{James Schneeloch}
\email{james.schneeloch.1@afrl.af.mil}
\affiliation{Air Force Research Laboratory, Information Directorate, Rome, New York, 13441, USA}

\date{\today}

\begin{abstract}
These comments contain two primary improvements to the aforementioned article:

Where previously mentioned in passing in section 3.4 of our article, we now provide, using the same basic steps as in Bennink's paper \cite{bennink2010optimal} (with some expanded discussion), a full derivation for the absolute generation rate of photon pairs from Spontaneous-Parametric Down-Conversion (SPDC) when using a focused gaussian pump beam as input, and collecting the output biphotons in a pair of focused gaussian collection modes. We do this to both show consistency between our formalism and his treatment, and to be able to use it to predict the absolute brightness of photon pair generation in these sources. In doing so, our formula differs from that in \cite{bennink2010optimal} by only a constant factor based on the ratio of products of indices of refraction.

Second, we correct an abuse of notation in our main work in which the quantization dimensions of the cavity modes of the electromagnetic field are conflated with the dimensions of the nonlinear medium. While this does not meaningfully affect the final formulas we obtain for the generation rates in the paper (adding a dimensionless correction factor of order unity), we use the gaussian-mode SPDC derivation here to show how the factors of the quantization dimensions disappear when converting from a discrete-momentum basis of rectangular cavity modes to a continuous frequency basis of gaussian spatial modes. In doing this, we find that applying this correction factor to our previous theoretical predictions for the photon pair generation rates brings them closer to the obtained experimental values.
\end{abstract}


\maketitle
\thispagestyle{fancy}

\newpage
This discourse is highly abbreviated, but with additional explanation where necessary. For context, we recommend reading this in parallel with our main work \cite{Schneeloch_2019_RateRef}.

\section{Absolute Brightness of SPDC for gaussian modes using Bennink derivation}
\subsection{The Nonlinear Hamiltonian}
From our main work \cite{Schneeloch_2019_RateRef}, the nonlinear Hamiltonian for Spontaneous Parametric Down-Conversion (SPDC) is given by:
\begin{equation}
\hat{H}_{NL}=\frac{1}{3}\int d^{3}r\;\big(\zeta_{ij\ell}^{(2)}(\vec{r})\hat{D}_{i}^{+}(\vec{r},t)\hat{D}_{j}^{-}(\vec{r},t)\hat{D}_{\ell}^{-}(\vec{r},t) + H.c.,\big),
\end{equation}
where repeated coordinate indices are summed over according to the Einstein summation convention. Here: $\zeta_{ij\ell}^{(2)}(\vec{r})$ is the second-order inverse optical susceptibility tensor of the medium the light is interacting with, and $\hat{D}_{i}^{+}(\vec{r},t)$ is the positive frequency component of the electric displacement field operator, such that:
\begin{equation}\label{equantexp}
\hat{D}^{+}(\vec{r},t)=\sum_{\vec{k},s}i\;\sqrt{\frac{\epsilon_{0}n^{2}_{\vec{k}}\hbar\omega_{\vec{k}}}{2 V}}\hat{a}_{\vec{k},s}(t)\vec{\epsilon}_{k,s}e^{i\vec{k}\cdot\vec{r}}e^{-i\omega_{k} t}.
\end{equation}
In this expression: $\epsilon_{0}$ is the permittivity of free space; $n_{\vec{k}}$ is the index of refraction for a photon of momentum $\vec{p}=\hbar \vec{k}$, where we use $\vec{k}$ to index momentum to eliminate extra factors of $\hbar=h/2\pi$, the quantum of angular momentum. With this, we point out that the angular frequency of the photon $\omega_{k}$ is also indexed by its momentum. In addition: $\hat{a}_{\vec{k},s}$ is the annihilation operator of a photon with momentum indexed by $\vec{k}$ and polarization indexed by $s$, and $\vec{\epsilon}_{k,s}$ is a unit vector in the direction of that polarization. The field modes are initially defined with respect to a rectangular cavity with volume $V=\mathcal{L}_{x}\mathcal{L}_{y}\mathcal{L}_{z}$, but this changes as we change basis. To distinguish the dimensions of the quantization volume from the dimensions of the nonlinear medium, we let $\{L_{x},L_{y},L_{z}\}$ be the dimensions of the nonlinear medium.

\subsubsection{Inverse optical susceptibility}
Where orders of the ordinary optical susceptibility $\chi^{(n)}$ are defined based on the power series expansion of the displacement field $\vec{D}$ in powers of the electric field $\vec{E}$:
\begin{equation}\label{DpowerSeries}
D_{i}= \epsilon_{0}\big[(\delta_{ij} + \chi_{ij}^{(1)})E_{j} + \chi_{ijk}^{(2)}E_{j}E_{k} + \chi_{ijk\ell}^{(3)}E_{j}E_{k}E_{\ell}+...  \big],
\end{equation}
orders of the \emph{inverse} optical susceptibility $\zeta^{(n)}$ are defined based on the power series expansion of $\vec{E}$ in powers of $\vec{D}$:
\begin{equation}\label{EPowerSeries}
E_{i}= \zeta_{ij}^{(1)}D_{j} + \zeta_{ijk}^{(2)}D_{j}D_{k} + \zeta_{ijk\ell}^{(3)}D_{j}D_{k}D_{\ell}+...  
\end{equation}
By substituting the power series for $\vec{E}$  \eqref{EPowerSeries} into the terms of expression for the power series of $\vec{D}$ \eqref{DpowerSeries}, one can obtain relations between forward and inverse susceptibilities. In doing this, the linear term gives us the relation between first order (forward and inverse) susceptibilities, and the quadratic term gives us the relation between second order susceptibilities. From \cite{quesada2017you}, we have the relations:
\begin{subequations}
\begin{align}
\epsilon_{0}\zeta_{ij}^{(1)} &= (\delta +\chi ^{(1)})^{-1}_{ij},\\
\zeta_{ijk}^{(2)}&=-\epsilon_{0}\zeta_{i\ell}^{(1)}\chi_{\ell mn}^{(2)}\zeta^{(1)}_{mj}\zeta^{(1)}_{nk}.
\end{align}
\end{subequations}
Since these relations come from an expression of $\vec{D}$ in powers of $\vec{D}$, the minus sign in $\zeta_{ijk}^{(2)}$ comes from the fact that the quadratic terms must add up to zero.

Assuming polarizations have already been set for the pump, signal, and idler fields (by phase matching or by design), we can directly express the effective inverse susceptibility in terms of the effective susceptibility. Where the refractive index $n$ is related to the first-order susceptibility $\chi^{(1)}_{\text{eff}}$ by the relation:
\begin{equation}
n^{2}=1+\chi^{(1)}_{\text{eff}},
\end{equation}
we have the approximate formula:
\begin{equation}\label{Inversechi2Formula}
\zeta_{\text{eff}}^{(2)}=-\frac{\chi_{\text{eff}}^{(2)}}{\epsilon_{0}^{2}n_{p}^{2}n_{1}^{2}n_{2}^{2}}.
\end{equation}

Note: To get this formula specified at the appropriate frequencies, we can use the approximate relations:

\begin{subequations}
\begin{align}
D(\omega_{p})&=\epsilon_{0}(1+\chi^{(1)}(\omega_{p}))E(\omega_{p}) + \epsilon_{0}\chi^{2}(\omega_{p}=\omega_{1}+\omega_{2})E(\omega_{1})E(\omega_{2})\\
E(\omega_{p})&= \zeta^{(1)}(\omega_{p})D(\omega_{p}) +\zeta^{(2)}(\omega_{p}=\omega_{1}+\omega_{2})D(\omega_{1})D(\omega_{2})\\
E(\omega_{1})&= \zeta^{(1)}(\omega_{1})D(\omega_{1}) +\zeta^{(2)}(\omega_{1}=\omega_{p}-\omega_{2})D(\omega_{p})D(\omega_{2}) \\
E(\omega_{2})&= \zeta^{(1)}(\omega_{2})D(\omega_{2}) +\zeta^{(2)}(\omega_{2}=\omega_{p}-\omega_{1})D(\omega_{3})D(\omega_{1}) 
\end{align}
\end{subequations}
and perform the same substitutions as before.

\subsubsection{Converting to Gaussian beam modes}
In general, we can express a creation operator in a new basis as a linear combination of operators in a previous basis through a simple linear transformation given by matrix $\tilde{C}_{\vec{q},\vec{\mu}}$:
\begin{equation}
\hat{a}^{\dagger}_{(\vec{q},k_{z},s)}=\sum_{\vec{\mu}}\tilde{C}_{\vec{q},\vec{\mu}}\;\hat{a}^{\dagger}_{(\vec{\mu},k_{z},s)}.
\end{equation}
Here, $\vec{q}$ is the projection of the momentum $\vec{k}$ onto the transverse plane (i.e., $\vec{q}=(k_{x},k_{y},0)$).

Preservation of the commutation relations $[\hat{a}_{(\vec{q},k_{z},s)},\hat{a}^{\dagger}_{(\vec{q}',k_{z},s)}]=\delta_{\vec{q},\vec{q}'}$ between bases implies $\tilde{C}_{\vec{q},\vec{\mu}}$ is unitary:
\begin{equation}
\delta_{\vec{q},\vec{q}'}=\sum_{\vec{\mu}}\tilde{C}^{*}_{\vec{q},\vec{\mu}}\tilde{C}_{\vec{q'},\vec{\mu}}.
\end{equation}

Converting to the transverse mode basis, we define the transverse spatial mode function  $g_{\vec{\mu}}(x,y)$ by the relation:
\begin{equation}
\sum_{\vec{q}}\frac{\tilde{C}_{\vec{q},\vec{\mu}}}{\sqrt{\mathcal{L}_{x}\mathcal{L}_{y}}}\;e^{-i\vec{q}\cdot\vec{r}}= \; g_{\vec{\mu}}(x,y).
\end{equation}

This definition (approximately) makes $g_{\vec{\mu}}(x,y)$ square-integrable and normalized:
\begin{align}
\int dxdy |g_{\mu}(x,y)|^{2} &= \sum_{\vec{q}}\sum_{\vec{q}'}\int dxdy \frac{\tilde{C}_{\vec{q},\vec{\mu}}\tilde{C}^{*}_{\vec{q}',\vec{\mu}}}{\mathcal{L}_{x}\mathcal{L}_{y}} e^{-i (\vec{q}-\vec{q}')\cdot \vec{r}}\nn\\
&= \sum_{\vec{q}}\sum_{\vec{q}'}\frac{\tilde{C}_{\vec{q},\vec{\mu}}\tilde{C}^{*}_{\vec{q}',\vec{\mu}}}{\mathcal{L}_{x}\mathcal{L}_{y}} (2\pi)^{2}\delta(\vec{q}-\vec{q}')\nn\\
&\approx \sum_{\vec{q}}\frac{\mathcal{L}_{x}\mathcal{L}_{y}}{(2\pi)^{2}}\int dq_{x}'dq_{y}'\frac{\tilde{C}_{\vec{q},\vec{\mu}}\tilde{C}^{*}_{\vec{q}',\vec{\mu}}}{\mathcal{L}_{x}\mathcal{L}_{y}} (2\pi)^{2}\delta(\vec{q}-\vec{q}')\nn\\
&=\sum_{\vec{q}}|\tilde{C}_{\vec{q},\vec{\mu}}|^{2} = 1
\end{align}
Here, the approximation is that the sum over discrete transverse momentum can be approximated as an integral over continuous transverse momentum.

In the transverse mode basis, the displacement field operator is then given by:
\begin{equation}
\hat{D}^{+}(\vec{r},t)=i\!\!\sum_{\vec{\mu},k_{z},s}\!\!\sqrt{\frac{\epsilon_{0}n^{2}_{k_{z}}\!\hbar\omega_{k_{z}}}{2 \mathcal{L}_{z}}}\vec{\epsilon}_{k_{z},s}g_{\vec{\mu}}(x,y)e^{i k_{z} z}e^{-i\omega t}\hat{a}_{\vec{\mu},k_{z},s}.
\end{equation}

\subsubsection{Simplifying the discrete Hamiltonian}
In this new basis of transverse modes, and concerning ourselves with just one triplet of spatial modes $(\vec{\mu}_{p},\vec{\mu_{1}},\vec{\mu}_{2})$ the Hamiltonian takes the form:
\begin{align}
\hat{H}_{NL}&=\frac{1}{3}\int d^{3}r\;\big(\zeta_{ij\ell}^{(2)}(\vec{r})\nn\\
&\cdot i\!\!\sum_{k_{pz},sp}\!\!\sqrt{\frac{\epsilon_{0}n^{2}_{p}\!\hbar\omega_{p}}{2 \mathcal{L}_{z}}}(\vec{\epsilon}_{k_{pz},sp})_{i}g_{\vec{\mu}_{p}}(x,y)e^{i k_{pz} z}e^{-i\omega_{p} t}\hat{a}_{\vec{\mu}_{p},k_{pz},sp}\nn\\
&\cdot -i\!\!\sum_{k_{1z},s1}\!\!\sqrt{\frac{\epsilon_{0}n^{2}_{1}\!\hbar\omega_{1}}{2 \mathcal{L}_{z}}}(\vec{\epsilon}_{k_{1z},s1})_{j}g_{\vec{\mu}_{1}}^{*}(x,y)e^{-i k_{1z} z}e^{i\omega_{1} t}\hat{a}_{\vec{\mu}_{1},k_{1z},s1}^{\dagger}\nn\\
&\cdot -i\!\!\sum_{k_{2z},s2}\!\!\sqrt{\frac{\epsilon_{0}n^{2}_{2}\!\hbar\omega_{2}}{2 \mathcal{L}_{z}}}(\vec{\epsilon}_{k_{2z},s2})_{\ell}g_{\vec{\mu}_{2}}^{*}(x,y)e^{-i k_{2z} z}e^{i\omega_{2} t}\hat{a}_{\vec{\mu}_{2},k_{2z},s2}^{\dagger}\nn\\
& \qquad + H.c.,\big).
\end{align}

Consolidating terms, we simplify the Hamiltonian:
\begin{align}
\hat{H}_{NL}&=\frac{-i}{3}\sum_{k_{pz},sp}\sum_{k_{1z},s1}\sum_{k_{2z},s2}\sqrt{\frac{\epsilon_{0}n^{2}_{p}\!\hbar\omega_{p}}{2 \mathcal{L}_{z}}}\sqrt{\frac{\epsilon_{0}n^{2}_{1}\!\hbar\omega_{1}}{2 L_{z}}}\sqrt{\frac{\epsilon_{0}n^{2}_{2}\!\hbar\omega_{2}}{2 \mathcal{L}_{z}}}\nn\\
&\cdot \int d^{3}r\;\big(\zeta_{ij\ell}^{(2)}(\vec{r})(\vec{\epsilon}_{k_{pz},sp})_{i}(\vec{\epsilon}_{k_{1z},s1})_{j}(\vec{\epsilon}_{k_{2z},s2})_{\ell}g_{\vec{\mu}_{p}}(x,y)g_{\vec{\mu}_{1}}^{*}(x,y)g_{\vec{\mu}_{2}}^{*}(x,y)\nn\\
&\cdot e^{i k_{pz} z}e^{-i k_{1z} z}e^{-i k_{2z} z}\big)e^{-i\omega_{p} t}e^{i\omega_{1} t}e^{i\omega_{2} t}\nn\\
&\cdot \hat{a}_{\vec{\mu}_{p},k_{pz},sp}\hat{a}_{\vec{\mu}_{1},k_{1z},s1}^{\dagger}\hat{a}_{\vec{\mu}_{2},k_{2z},s2}^{\dagger} + H.c.,\big).
\end{align}

We assume we are using one polarization of the pump, signal, and idler fields, and carry out the summation over the components of the inverse nonlinear susceptibility $\zeta^{(2)}$. We also define $\Delta k_{z} \equiv k_{1z}+k_{2z}-k_{pz}$ and $\Delta\omega \equiv \omega_{1}+\omega_{2}-\omega_{p}$. Finally, we assume the pump is bright enough to be treated as an undepleted classical field, and replace its annihilation operator with a corresponding coherent state amplitude:
\begin{align}
\hat{H}_{NL}&=-2i\sum_{k_{pz}}\sum_{k_{1z}}\sum_{k_{2z}}\sqrt{\frac{\epsilon_{0}^{3}n^{2}_{p}n^{2}_{1}n^{2}_{2}\hbar^{3}\omega_{p}\omega_{1}\omega_{2}}{8 \mathcal{L}_{z}^{3}}}\nn\\
&\cdot \int d^{3}r\;\big(\zeta_{\text{eff}}^{(2)}(\vec{r})g_{\vec{\mu}_{p}}(x,y)g_{\vec{\mu}_{1}}^{*}(x,y)g_{\vec{\mu}_{2}}^{*}(x,y)e^{-i\Delta k_{z} z}\big)\nn\\
&\cdot e^{i\Delta\omega t}\cdot \alpha_{\vec{\mu}_{p},k_{pz}}\hat{a}_{\vec{\mu}_{1},k_{1z}}^{\dagger}\hat{a}_{\vec{\mu}_{2},k_{2z}}^{\dagger}+ H.c.,\big).
\end{align}

The nonlinear susceptibility $\chi^{(2)}_{\text{eff}}$ is more commonly used than its inverse $\zeta_{\text{eff}}^{(2)}$, so we make the substitution with the relation \eqref{Inversechi2Formula}. In addition, 
we assume an isotropic nonlinear medium (not counting periodic poling) to make the substitution:
\begin{equation}
\chi_{\text{eff}}^{(2)}=2\bar{\chi}_{\text{eff}}^{(2)}d_{\text{eff}}.
\end{equation}
Here, $\bar{\chi}_{\text{eff}}^{(2)}$ is a function that is unity inside the nonlinear medium, and zero outside (or alternating between $+1$ and $-1$ inside the medium in the case of periodic poling). $d_{\text{eff}}$ is the bulk effective nonlinear susceptibility used to compare different nonlinear materials. This gives us the hamiltonian:
\begin{align}
\hat{H}_{NL}&=i \sum_{k_{pz}}\sum_{k_{1z}}\sum_{k_{2z}}2 d_{\text{eff}}\sqrt{\frac{\hbar^{3}\omega_{p}\omega_{1}\omega_{2}}{2 \epsilon_{0} \mathcal{L}_{z}^{3}n^{2}_{p}n^{2}_{1}n^{2}_{2}}}\nn\\
&\qquad\cdot \int d^{3}r\;\big(\bar{\chi}_{\text{eff}}^{(2)}(\vec{r})g_{\vec{\mu}_{p}}(x,y)g_{\vec{\mu}_{1}}^{*}(x,y)g_{\vec{\mu}_{2}}^{*}(x,y)e^{-i\Delta k_{z} z}\big)\nn\\
&\qquad\qquad\cdot e^{i\Delta\omega t}\cdot \alpha_{\vec{\mu}_{p},k_{pz}}\hat{a}_{\vec{\mu}_{1},k_{1z}}^{\dagger}\hat{a}_{\vec{\mu}_{2},k_{2z}}^{\dagger}+ H.c.
\end{align}

Next, we define the overlap integral:
\begin{equation}
\Phi(\Delta k_{z})\equiv\int d^{3}r\;\big(\bar{\chi}_{\text{eff}}^{(2)}(\vec{r})g_{\vec{\mu}_{p}}(x,y)g_{\vec{\mu}_{1}}^{*}(x,y)g_{\vec{\mu}_{2}}^{*}(x,y)e^{-i\Delta k_{z} z}\big),
\end{equation}
so that the Hamiltonian simplifies to:
\begin{align}
\hat{H}_{NL}&=i \sum_{k_{pz}}\sum_{k_{1z}}\sum_{k_{2z}}2 d_{\text{eff}}\sqrt{\frac{\hbar^{3}\omega_{p}\omega_{1}\omega_{2}}{2 \epsilon_{0} \mathcal{L}_{z}^{3}n^{2}_{p}n^{2}_{1}n^{2}_{2}}}\alpha_{\vec{\mu}_{p},k_{pz}}\Phi(\Delta k_{z})e^{i\Delta\omega t}\cdot \hat{a}_{\vec{\mu}_{1},k_{1z}}^{\dagger}\hat{a}_{\vec{\mu}_{2},k_{2z}}^{\dagger}+ H.c.
\end{align}

\subsubsection{Transformation to continuous frequencies and  eliminating quantization dimensions}
Here, we convert the hamiltonian from sums over discrete momenta into integrals over continuous frequency.
In particular, we must maintain the relation:
\begin{equation}
\sum_{k}\hat{a}_{k}^{\dagger}\hat{a}_{k} \approx \int dk \hat{a}^{\dagger}(k)\hat{a}(k)\approx \int d\omega \hat{a}^{\dagger}(\omega)\hat{a}(\omega).
\end{equation}
From here, it is clear that $\hat{a}_{k}$ and $\hat{a}(k)$ (for discrete and continuous momentum, respectively) have different dimensions, and are related differently to one another as well as to $\hat{a}(\omega)$. Indeed, we find between continuous frequency and discrete momentum, that:
\begin{equation}
\hat{a}^{\dagger}(\omega) \approx \sqrt{\frac{\mathcal{L}_{z}n_{g}}{2\pi c}} \hat{a}_{k}^{\dagger}
\end{equation}
where $n_{g}$ is the group index of refraction.

For a general function, we have:
\begin{equation}
\sum_{k_{1z},k_{2z}}\approx\Big(\frac{\mathcal{L}_{z}}{2\pi}\Big)^{2}\frac{n_{g1}n_{g2}}{c^{2}}\int d\omega_{1}d\omega_{2},
\end{equation}
so that for the sum:
\begin{align}
\sum_{k_{pz}}\sum_{k_{1z}}\sum_{k_{2z}}&\alpha_{\vec{\mu}_{p},k_{pz}}\hat{a}_{\vec{\mu}_{1},k_{1z}}^{\dagger}\hat{a}_{\vec{\mu}_{2},k_{2z}}^{\dagger} \rightarrow \nn\\
&\rightarrow\Big(\frac{\mathcal{L}_{z}}{2\pi}\Big)^{3}\frac{n_{g1}n_{g2}n_{gp}}{c^{3}} \Big(\sqrt{\frac{\mathcal{L}_{z}n_{gp}}{2\pi c}}\sqrt{\frac{\mathcal{L}_{z}n_{g1}}{2\pi c}}\sqrt{\frac{\mathcal{L}_{z}n_{g2}}{2\pi c}}\Big)^{-1}\int d\omega_{p}d\omega_{1}d\omega_{2} \alpha(\omega_{p})\hat{a}^{\dagger}_{\vec{\mu}_{1}}(\omega_{1})\hat{a}^{\dagger}_{\vec{\mu}_{2}}(\omega_{2}),
\end{align}
which gives us:
\begin{equation}
\sum_{k_{pz}}\sum_{k_{1z}}\sum_{k_{2z}}\alpha_{\vec{\mu}_{p},k_{pz}}\hat{a}_{\vec{\mu}_{1},k_{1z}}^{\dagger}\hat{a}_{\vec{\mu}_{2},k_{2z}}^{\dagger} \rightarrow 
\Big(\frac{\mathcal{L}_{z}}{2\pi}\Big)^{3/2}\sqrt{\frac{n_{g1}n_{g2}n_{gp}}{c^{3}}} \int d\omega_{p}d\omega_{1}d\omega_{2} \alpha(\omega_{p})\hat{a}^{\dagger}_{\vec{\mu}_{1}}(\omega_{1})\hat{a}^{\dagger}_{\vec{\mu}_{2}}(\omega_{2}).
\end{equation}

With this, the Hamiltonian becomes:
\begin{align}
\hat{H}_{\!N\!L}\!\!&=\!i 2 d_{\text{eff}}\!\Big(\!\frac{\mathcal{L}_{z}}{2\pi}\!\Big)^{3/2}\!\!\!\!\sqrt{\!\frac{n_{g1}n_{g2}n_{gp}}{c^{3}}}\!\!\!\int \!\! d\omega_{p}d\omega_{1}d\omega_{2}\sqrt{\!\frac{\hbar^{3}\omega_{p}\omega_{1}\omega_{2}}{2 \epsilon_{0} \mathcal{L}_{z}^{3}n^{2}_{p}n^{2}_{1}n^{2}_{2}}}\alpha(\omega_{p})\Phi(\Delta k_{z})e^{i\Delta\omega t}\!\!\cdot\! \hat{a}^{\dagger}_{\vec{\mu}_{1}}\!(\omega_{1})\hat{a}^{\dagger}_{\vec{\mu}_{2}}\!(\omega_{2})\!+\! H.c.
\end{align}
which simplifies to:
\begin{equation}
\boxed{\hat{H}_{NL}=i\hbar \sqrt{\frac{\hbar d_{\text{eff}}^{2}}{4 \pi^{3} \epsilon_{0} c^{3}}}\sqrt{\frac{n_{g1}n_{g2}n_{gp}}{n^{2}_{p}n^{2}_{1}n^{2}_{2}}}\!\!\int\!\! d\omega_{p}d\omega_{1}d\omega_{2}\sqrt{\omega_{p}\omega_{1}\omega_{2}}\alpha(\omega_{p})\Phi(\Delta k_{z})e^{i\Delta\omega t}\cdot \hat{a}^{\dagger}_{\vec{\mu}_{1}}(\omega_{1})\hat{a}^{\dagger}_{\vec{\mu}_{2}}(\omega_{2})+ H.c.}
\end{equation}
In this simplified Hamiltonian, we see no factors of the quantization length $\mathcal{L}_{z}$ (or of other quantization dimensions). Indeed, by both transforming to a basis of spatial modes and taking the continuum limit, we have incorporated the effects of a quantization volume (where necessary) into the definition of the spatial mode functions. This was derived on the assumption that the nonlinear medium has no reflectivity at its ends (e.g., by using antireflective coatings), as it would otherwise be in a cavity. However, for cavities with nonzero loss, the Hamiltonian will still look the same. What changes is that care must be taken to convert between the creation operators for light propagating inside the cavity and the the creation operators for photons propagating outside the cavity. These conversion factors account for the effects of resonance on the light that ultimately leaves the cavity.

\subsection{Finding the Down-converted state and pair generation rate}
With the Hamiltonian for SPDC fully simplified, we apply the first-order time evolution operator (from first-order perturbation theory) to obtain the state of the down-converted light. The initial state of the field is a coherent state in the pump frequency band, and the vacuum state at the signal/idler frequency bands, so that the final state is given by:
\begin{subequations}
\begin{align}
|\Psi_{\text{final}}\rangle &\approx \bigg( \mathbf{I} -\frac{i}{\hbar}\int_{-\infty}^{\infty} \!\!dt' \hat{H}_{NL}(t') \bigg)|\alpha_{p},0,0\rangle\\
&= |\alpha_{p},0,0\rangle + |\Psi_{\text{SPDC}}\rangle,
\end{align}
\end{subequations}
where the (unnormalized) biphoton state $|\Psi_{\text{SPDC}}\rangle$ is given by:
\begin{equation}
|\Psi_{\text{SPDC}}\rangle \!=\!\sqrt{\!\frac{\hbar d_{\text{eff}}^{2}}{\pi \epsilon_{0} c^{3}}}\sqrt{\frac{n_{g1}n_{g2}n_{gp}}{n^{2}_{p}n^{2}_{1}n^{2}_{2}}}\!\!\int\!\! d\omega_{p}d\omega_{1}d\omega_{2}\delta(\Delta\omega)\sqrt{\omega_{p}\omega_{1}\omega_{2}}\alpha(\omega_{p})\Phi(\Delta k_{z})\cdot \hat{a}^{\dagger}_{\vec{\mu}_{1}}(\omega_{1})\hat{a}^{\dagger}_{\vec{\mu}_{2}}(\omega_{2})|\alpha_{p},0,0\rangle.
\end{equation}
Next we separate out the pump photon number $N_{p}$ from its overall amplitude $\alpha(\omega_{p})$ and spectral amplitude $s(\omega_{p})$:
\begin{equation}
\alpha(\omega_{p})=s(\omega_{p})\sqrt{N_{p}},
\end{equation}
which gives us:
\begin{equation}
|\Psi_{\text{SPDC}}\rangle \!=\!\sqrt{\!\frac{\hbar N_{p} d_{\text{eff}}^{2}}{\pi \epsilon_{0} c^{3}}}\!\!\sqrt{\frac{n_{g1}n_{g2}n_{gp}}{n^{2}_{p}n^{2}_{1}n^{2}_{2}}}\!\!\int\!\! d\omega_{p}d\omega_{1}d\omega_{2}\delta(\!\Delta\omega)\sqrt{\omega_{p}\omega_{1}\omega_{2}}s(\omega_{p})\Phi(\Delta k_{z})\cdot \hat{a}^{\dagger}_{\vec{\mu}_{1}}(\omega_{1})\hat{a}^{\dagger}_{\vec{\mu}_{2}}(\omega_{2})|\alpha_{p},0,0\rangle.
\end{equation}

Assuming the variation in $\sqrt{\omega_{p}\omega_{1}\omega_{2}}$ is much slower than the other terms in the integral, we approximate it with its value at the central frequencies $\sqrt{\omega_{p0}\omega_{10}\omega_{20}}$, and simplify by taking it outside the integral:
\begin{equation}
|\Psi_{\text{SPDC}}\rangle \approx\!\!\sqrt{\frac{\hbar N_{p} d_{\text{eff}}^{2}}{\pi \epsilon_{0} c^{3}}}\!\!\sqrt{\frac{n_{g1}n_{g2}n_{gp}}{n^{2}_{p}n^{2}_{1}n^{2}_{2}}}\!\!\sqrt{\omega_{p0}\omega_{10}\omega_{20}}\!\!\int\!\! d\omega_{p}d\omega_{1}d\omega_{2}\delta(\!\Delta\omega)s(\omega_{p})\Phi(\Delta k_{z})\cdot \hat{a}^{\dagger}_{\vec{\mu}_{1}}(\omega_{1})\hat{a}^{\dagger}_{\vec{\mu}_{2}}(\omega_{2})|\alpha_{p},0,0\rangle.
\end{equation}
To isolate the state of the generated biphoton, we trace over the state of the pump field, which is independent in this approximation. Carrying out the integral over $\omega_{p}$ with $\delta(\Delta\omega)$ present in the integrand enforces energy conservation, setting $\omega_{p}=\omega_{1}+\omega_{2}$:
\begin{equation}
|\Psi_{\text{SPDC}}\rangle =\sqrt{\frac{\hbar N_{p} d_{\text{eff}}^{2}}{\pi \epsilon_{0} c^{3}}}\sqrt{\frac{n_{g1}n_{g2}n_{gp}}{n^{2}_{p}n^{2}_{1}n^{2}_{2}}}\sqrt{\omega_{p0}\omega_{10}\omega_{20}}\int d\omega_{1}d\omega_{2}s(\omega_{1}+\omega_{2})\Phi(\Delta k_{z})\cdot \hat{a}^{\dagger}_{\vec{\mu}_{1}}(\omega_{1})\hat{a}^{\dagger}_{\vec{\mu}_{2}}(\omega_{2})|0,0\rangle.
\end{equation}

Defining vacuum wavelengths $\lambda$ in terms of angular frequency $\omega$:
\begin{equation}
\omega = \frac{2\pi c}{\lambda},
\end{equation}
we express $|\Psi_{\text{SPDC}}\rangle$ in terms of wavelengths (noting the effect of refractive index):
\begin{equation}
|\Psi_{\text{SPDC}}\rangle =\sqrt{\frac{\hbar N_{p} d_{\text{eff}}^{2}}{\pi \epsilon_{0} c^{3}}}\sqrt{\frac{n_{g1}n_{g2}n_{gp}}{n_{p}^{2}n_{1}^{2}n_{2}^{2}}}\sqrt{\frac{8\pi^{3}c^{3}}{\lambda_{p0}\lambda_{10}\lambda_{20}}}\int d\omega_{1}d\omega_{2}s(\omega_{1}+\omega_{2})\Phi(\Delta k_{z})\cdot \hat{a}^{\dagger}_{\vec{\mu}_{1}}(\omega_{1})\hat{a}^{\dagger}_{\vec{\mu}_{2}}(\omega_{2})|0,0\rangle,
\end{equation}
which simplifies to:
\begin{equation}
|\Psi_{\text{SPDC}}\rangle =(2 d_{\text{eff}})\sqrt{\frac{2\pi^{2}\hbar N_{p}}{\epsilon_{0}\lambda_{p0}\lambda_{10}\lambda_{20}}}\sqrt{\frac{n_{g1}n_{g2}n_{gp}}{n_{p}^{2}n_{1}^{2}n_{2}^{2}}}\int d\omega_{1}d\omega_{2}s(\omega_{1}+\omega_{2})\Phi(\Delta k_{z})\cdot \hat{a}^{\dagger}_{\vec{\mu}_{1}}(\omega_{1})\hat{a}^{\dagger}_{\vec{\mu}_{2}}(\omega_{2})|0,0\rangle.
\end{equation}

In Bennink's Reference \cite{bennink2010optimal}, we have that their overlap integral denoted by $\mathscr{O}(\omega_{1},\omega_{2})$ is given by the relation $\mathscr{O}(\omega_{1},\omega_{2})=2 d_{\text{eff}} \Phi(\Delta k_{z})$, so that from our derivation, the state of the down-converted light can be expressed as:
\begin{equation}
|\Psi_{\text{SPDC}}\rangle =\sqrt{\frac{2\pi^{2}\hbar N_{p}}{\epsilon_{0}\lambda_{p0}\lambda_{10}\lambda_{20}}}\sqrt{\frac{n_{g1}n_{g2}n_{gp}}{n_{p}^{2}n_{1}^{2}n_{2}^{2}}}\int d\omega_{1}d\omega_{2}s(\omega_{1}+\omega_{2})\mathscr{O}(\omega_{1},\omega_{2})\cdot \hat{a}^{\dagger}_{\vec{\mu}_{1}}(\omega_{1})\hat{a}^{\dagger}_{\vec{\mu}_{2}}(\omega_{2})|0,0\rangle.
\end{equation}

Defining the joint spectral amplitude (JSA) $\psi(\omega_{1},\omega_{2})$ to incorporate all coefficients:
\begin{equation}
\psi(\omega_{1},\omega_{2})=\sqrt{\frac{2\pi^{2}\hbar N_{p}}{\epsilon_{0}\lambda_{p0}\lambda_{10}\lambda_{20}}}\sqrt{\frac{n_{g1}n_{g2}n_{gp}}{n_{p}^{2}n_{1}^{2}n_{2}^{2}}}s(\omega_{1}+\omega_{2})\mathscr{O}(\omega_{1},\omega_{2}),
\end{equation}
we have for the state:
\begin{equation}
|\Psi_{\text{SPDC}}\rangle =\int d\omega_{1}d\omega_{2} \;\psi(\omega_{1},\omega_{2})\cdot \hat{a}^{\dagger}_{\vec{\mu}_{1}}(\omega_{1})\hat{a}^{\dagger}_{\vec{\mu}_{2}}(\omega_{2})|0,0\rangle.
\end{equation}
Since to first order, the biphoton state has amplitudes only for either the vacuum state or that of a single biphoton, the average number of generated photon pairs per pump pulse (which has on average $N_{p}$ photons) is given by the probability integral:
\begin{equation}
\langle N_{pairs}\rangle = \int d\omega_{1}d\omega_{2}|\psi(\omega_{1},\omega_{2})|^{2}.
\end{equation}
From this, we can take the \emph{rate} of generated photon pairs simply by substituting for $N_{p}$ the rate of pump photons passing through the nonlinear medium (i.e., pump power $P$ divided by $\hbar\omega_{p}$).

\subsection{The Bennink Overlap Integral of Gaussian-beam SPDC}
Up until this point, we have made no assumptions about the form of the transverse mode function $g_{\vec{\mu}}(x,y)$. Here, and through the rest of this work, we will take it to be the zero-order Gaussian beam mode given by:
\begin{equation}
g_{\mu}(x,y) = \sqrt{\frac{k z_{R}}{\pi}}\frac{1}{q}e^{-ik\frac{x^{2}+y^{2}}{2q}},
\end{equation}
where the complex beam parameter $q=z+ i z_{R}$ and the Rayleigh range $z_{R}=\frac{\pi w_{0}^{2} n}{\lambda}=\frac{k w_{0}^{2}}{2}$. Since we are not indexing over multiple transverse modes, we eliminate the vector from the subscript $\mu$ for simplicity. In addition we define $|\vec{k}|$ as $k$ to condense notation.

Here, we define the beam focal parameter $\xi_{j}$ where index $j=\{p,1,2\}$ denotes pump, signal, or idler beams, respectively:
\begin{equation}
\xi_{j} = \frac{L_{z}}{k_{j}w_{j}^{2}}=\frac{L_{z}}{2 z_{R}}.
\end{equation}

Now we look at the product of the Gaussian modes in the overlap integral for further simplification:
\begin{equation}
g_{\mu_{p}}(x,y)g_{\mu_{1}}^{*}(x,y)g_{\mu_{2}}^{*}(x,y) = \frac{w_{p}w_{1}w_{2}}{(\pi/2)^{3/2}}\frac{k_{p}}{2q_{p}}\frac{k_{1}}{2q_{1}^{*}}\frac{k_{2}}{2q_{2}^{*}}e^{-ik_{p}\frac{x^{2}+y^{2}}{2q_{p}}}e^{ik_{1}\frac{x^{2}+y^{2}}{2q_{1}^{*}}}e^{ik_{2}\frac{x^{2}+y^{2}}{2q_{2}^{*}}}.
\end{equation}

Let us define a scaled beam parameter $\bar{q}$:
\begin{equation}
\bar{q}= \frac{2i}{k}q=\frac{2i}{k}(z + i z_{R})=-w^{2}+\frac{2i}{k}z.
\end{equation}
This simplifies the product of Gaussian modes into:
\begin{align}
g_{\mu_{p}}(x,y)g_{\mu_{1}}^{*}(x,y)g_{\mu_{2}}^{*}(x,y) &= \frac{w_{p}w_{1}w_{2}}{(\pi/2)^{3/2}}\frac{-i}{\bar{q}_{p}\bar{q}_{1}^{*}\bar{q}_{2}^{*}}e^{-(x^{2}+y^{2})\big(\frac{1}{\bar{q}_{p}}+\frac{1}{\bar{q}_{1}^{*}}+\frac{1}{\bar{q}_{2}^{*}}\big)}\nn\\
&= \frac{w_{p}w_{1}w_{2}}{(\pi/2)^{3/2}}\frac{-i}{\bar{q}_{p}\bar{q}_{1}^{*}\bar{q}_{2}^{*}}e^{-(x^{2}+y^{2})\big(\frac{\bar{q}_{p}\bar{q}_{1}^{*}+\bar{q}_{p}\bar{q}_{2}^{*}+\bar{q}_{1}^{*}\bar{q}_{2}^{*}}{\bar{q}_{p}\bar{q}_{1}^{*}\bar{q}_{2}^{*}}\big)}.
\end{align}
Integrating this product over $x$ and $y$, we have for the overlap integral:
\begin{equation}
\mathscr{O}(\omega_{1},\omega_{2})=2 d_{\text{eff}} \Phi(\Delta k_{z})=\int dz \chi_{\text{eff}}^{(2)}(z)\frac{w_{p}w_{1}w_{2}}{(\pi/2)^{3/2}}\frac{-i\pi}{\bar{q}_{p}\bar{q}_{1}^{*}\bar{q}_{2}^{*}}\big(\frac{\bar{q}_{p}\bar{q}_{1}^{*}+\bar{q}_{p}\bar{q}_{2}^{*}+\bar{q}_{1}^{*}\bar{q}_{2}^{*}}{\bar{q}_{p}\bar{q}_{1}^{*}\bar{q}_{2}^{*}}\big)^{-1}e^{-i\Delta k_{z} z},
\end{equation}
which simplifies to:
\begin{equation}
\boxed{\mathscr{O}(\omega_{1},\omega_{2})=-i\chi_{\text{eff}}^{(2)}\sqrt{\frac{8}{\pi}}w_{p}w_{1}w_{2}\int_{-L_{z}/2}^{L_{z}/2} dz\frac{e^{-i\Delta k_{z} z}}{\bar{q}_{p}\bar{q}_{1}^{*}+\bar{q}_{p}\bar{q}_{2}^{*}+\bar{q}_{1}^{*}\bar{q}_{2}^{*}}}.
\end{equation}
By approximating $\Delta k_{z}$ as $\Delta k=k_{1}+k_{2}-k_{p}$, we reproduce equation 9 in Bennink's paper \cite{bennink2010optimal} (up to a sign convention on $\Delta k$).

Now, we prepare the overlap integral to make it easier to evaluate the pair generation rate. As in \cite{bennink2010optimal}, we make the substitution: $2z/L_{z}=\ell$ so that $dz = \frac{L_{z}}{2}d\ell$:
\begin{equation}
\mathscr{O}(\omega_{1},\omega_{2})=-i\chi_{\text{eff}}^{(2)}\sqrt{\frac{8}{\pi}}w_{p}w_{1}w_{2}\frac{L_{z}}{2}\int_{-1}^{1} d\ell\frac{e^{-i\Delta k \frac{L_{z}\ell}{2}}}{\bar{q}_{p}\bar{q}_{1}^{*}+\bar{q}_{p}\bar{q}_{2}^{*}+\bar{q}_{1}^{*}\bar{q}_{2}^{*}},
\end{equation}
as well as the substitution: $\phi\equiv \Delta k L_{z}$
\begin{equation}
\mathscr{O}(\omega_{1},\omega_{2})=-i\chi_{\text{eff}}^{(2)}\sqrt{\frac{8}{\pi}}w_{p}w_{1}w_{2}\frac{L_{z}}{2}\int_{-1}^{1} d\ell\frac{e^{-i\phi\frac{\ell}{2}}}{\bar{q}_{p}\bar{q}_{1}^{*}+\bar{q}_{p}\bar{q}_{2}^{*}+\bar{q}_{1}^{*}\bar{q}_{2}^{*}}.
\end{equation}

As in \cite{bennink2010optimal}, we define the aggregate focal parameter $\xi$:
\begin{equation}
\xi \equiv \frac{k_{1}\xi_{1}(\xi_{2}-\xi_{p}) + k_{2}\xi_{2}(\xi_{1}-\xi_{p}) + k_{p}\xi_{p}(\xi_{1}+\xi_{2})}{k_{1}\xi_{1} + k_{2}\xi_{2} + k_{p}\xi_{p}},
\end{equation}
a quadratic coefficient $C$:
\begin{equation}
C\equiv \frac{(k_{p}-k_{1}-k_{2})\xi_{1}\xi_{2}\xi_{p}(k_{1}\xi_{1}+k_{2}\xi_{2}+k_{p}\xi_{p})}{(k_{1}\xi_{1}(\xi_{2}-\xi_{p}) + k_{2}\xi_{2}(\xi_{1}-\xi_{p}) + k_{p}\xi_{p}(\xi_{1}+\xi_{2}))^{2}},
\end{equation}
and a normalization coefficient $D$:
\begin{equation}
D\equiv \frac{k_{1}k_{2}k_{p}\xi_{1}\xi_{2}\xi_{p}}{L(k_{1}\xi_{1}+k_{2}\xi_{2}+k_{p}\xi_{p})}.
\end{equation}
From these, we can express the overlap integral in a greatly simplified form:
\begin{equation}
\mathscr{O}(\omega_{1},\omega_{2})=-i\chi_{\text{eff}}^{(2)}\sqrt{\frac{2}{\pi}}w_{p}w_{1}w_{2}D\int_{-1}^{1} d\ell\frac{e^{-i\phi\frac{\ell}{2}}}{1 + i\ell\xi - C\xi^{2}\ell^{2}}.
\end{equation}
Note that under the approximation that the phase matching is not affected much by the curvature of the phase fronts (valid except for extreme focusing) $C\approx 0$. With this simplification, we are ready to compute the pair generation rate with a Gaussian beam pump into the pair of gaussian beam collection modes.

As mentioned before, the number of generated biphotons given the input of $N_{p}$ pump photons is given by the integral:
 \begin{equation}
\langle N_{pairs}\rangle = \int d\omega_{1}d\omega_{2}|\psi(\omega_{1},\omega_{2})|^{2},
\end{equation}
such that
\begin{equation}
|\psi(\omega_{1},\omega_{2})|^{2}=\frac{2\pi^{2}\hbar N_{p}}{\epsilon_{0}\lambda_{p}\lambda_{1}\lambda_{2}}\frac{n_{g1}n_{g2}n_{gp}}{n_{p}^{2}n_{1}^{2}n_{2}^{2}}|s(\omega_{1}+\omega_{2})|^{2}|\mathscr{O}(\omega_{1},\omega_{2})|^{2}.
\end{equation}
Expanding this out, we get:
\begin{equation}
\langle N_{pairs}\rangle=\frac{4\pi\hbar N_{p}|\chi_{\text{eff}}^{(2)}|^{2}}{\epsilon_{0}\lambda_{p}\lambda_{1}\lambda_{2}}\frac{n_{g1}n_{g2}n_{gp}}{n_{p}^{2}n_{1}^{2}n_{2}^{2}}w_{p}^{2}w_{1}^{2}w_{2}^{2}D^{2}\int d\omega_{1}d\omega_{2}|s(\omega_{1}+\omega_{2})|^{2}\left|\int_{-1}^{1} d\ell\frac{e^{-i\phi\frac{\ell}{2}}}{1 + i\ell\xi - C\xi^{2}\ell^{2}}\right|^{2}.
\end{equation}
As expected, a check of dimensions gives $\langle N_{pairs}\rangle$ as dimensionless.

For simplification, we condense all terms preceding the integral into variable $C_{n}$:
\begin{equation}
\langle N_{pairs}\rangle=C_{n}\int d\omega_{1}d\omega_{2}|s(\omega_{1}+\omega_{2})|^{2}\left|\int_{-1}^{1} d\ell\frac{e^{-i\phi\frac{\ell}{2}}}{1 + i\ell\xi - C\xi^{2}\ell^{2}}\right|^{2}.
\end{equation}

Taking the approximation that the quadratic coefficient $C\approx 0$, we can simplify the integral over $\ell$:
\begin{equation}
\left|\int_{-1}^{1} d\ell\frac{e^{i\phi\frac{\ell}{2}}}{1 + i\ell\xi - C\xi^{2}\ell^{2}}\right|^{2}\approx\iint_{-1}^{1} d\ell d\ell' \frac{e^{-i\frac{\phi}{2}(\ell-\ell')}}{(1+i\ell \xi )(1-i\ell'\xi)},
\end{equation}
so that
\begin{equation}
\langle N_{pairs}\rangle=C_{n}\iint_{-1}^{1} d\ell d\ell'\int d\omega_{1}d\omega_{2}|s(\omega_{1}+\omega_{2})|^{2} \frac{e^{-i\frac{\phi}{2}(\ell-\ell')}}{(1+i\ell \xi )(1-i\ell'\xi)}.
\end{equation}

Here we define the phase matching function $\phi$ to first order in frequency difference:
\begin{equation}\label{phasematchingfunction}
\phi \approx \Big(\frac{n_{g1} + n_{g2}-2 n_{gp}}{2c}\delta\omega_{p} +\frac{n_{g1}-n_{g2}}{2c}\delta\omega_{-}\Big)L_{z},
\end{equation}
where $\delta \omega_{p}= (\omega_{1}-\omega_{10}) + (\omega_{2}-\omega_{20})$, and $\delta \omega_{-}= (\omega_{1}-\omega_{10}) - (\omega_{2}-\omega_{20})$. With this, the dependence of $\phi$ on $\delta\omega_{p}$ can be neglected for a sufficiently narrowband pump (i.e., such that $\phi$ is nearly constant as $\delta\omega_{P}$ is varied over the pump bandwidth). This form of the phase matching function works when the group indices of refraction at the signal and idler central frequencies are different enough that we need not consider higher order dispersion in describing the spectrum of the biphoton. For example, this phase matching works well for type-0 or type-I \emph{non}degenerate SPDC, but not for type-0 or type-I degenerate SPDC. For type-II SPDC, it works well for both the degenerate and nondegenerate phase matching cases.

To date, there is no companion treatment $\langle N_{pairs}\rangle$ for degenerate type-I  (or type-0) SPDC using this same formalism, where:
\begin{equation}
\phi\approx \left(\frac{1}{4}\kappa_{0}\delta\omega_{-}^{2}\right) L_{z},
\end{equation}
but comparable studies on the effect of pump focusing on type-I degenerate SPDC exist \cite{bernecker2022spatial,anwar2017selecting,guilbert2014enhancing,smirr2013optimal,PhysRevA.77.043834,PhysRevA.106.043719}. Among other reasons for this is that when $\phi$ is a quadratic function of $\delta\omega_{-}$, the integrals over $\ell$ and $\ell'$ cannot be computed in terms of elementary functions, even though the integrals over $\omega_{1}$ and $\omega_{2}$ can. However, where $\xi$ is small enough that the integral over $\phi$ may still be approximated as a delta function, we should expect the same qualitative behavior for the generation rate as our formula for type-I degenerate SPDC (equation 42 in \cite{Schneeloch_2019_RateRef}).

Using the phase-matching function \eqref{phasematchingfunction} for nondegenerate/type-II SPDC, we change variables in the integration for $\langle N_{pairs}\rangle $ according to the Jacobian:
\begin{equation}
d\omega_{1}d\omega_{2}=\frac{c}{|n_{g1}-n_{g2}|L_{z}}d\omega_{p}d\phi,
\end{equation}
where $\omega_{p}=\omega_{1}+\omega_{2}$,
which gives us
\begin{equation}
\langle N_{pairs}\rangle=C_{n}\frac{c}{|n_{g1}-n_{g2}|L_{z}}\iint_{-1}^{1} d\ell d\ell'\int d\omega_{p}d\phi |s(\omega_{p})|^{2} \frac{e^{-i\frac{\phi}{2}(\ell-\ell')}}{(1+i\ell \xi )(1-i\ell'\xi)}.
\end{equation}
The integral over $\phi$ is straightforward, becoming a delta function:
\begin{equation}
\langle N_{pairs}\rangle=C_{n}\frac{4\pi c}{|n_{g1}-n_{g2}|L_{z}}\iint_{-1}^{1} d\ell d\ell'\int d\omega_{p} |s(\omega_{p})|^{2} \frac{\delta(\ell-\ell')}{(1+i\ell \xi )(1-i\ell'\xi)}.
\end{equation}
The magnitude square of the normalized pump spectral amplitude integrates to unity:
\begin{equation}
\langle N_{pairs}\rangle=C_{n}\frac{4\pi c}{|n_{g1}-n_{g2}|L_{z}}\iint_{-1}^{1} d\ell d\ell'\frac{\delta(\ell-\ell')}{(1+i\ell \xi )(1-i\ell'\xi)},
\end{equation}
and we integrate over $\ell'$ to eliminate the delta function:
\begin{equation}
\langle N_{pairs}\rangle=C_{n}\frac{4\pi c}{|n_{g1}-n_{g2}|L_{z}}\int_{-1}^{1} d\ell \frac{1}{1+\ell^{2}\xi^{2}}.
\end{equation}

This function is straightforward to integrate:
\begin{equation}
\int_{-1}^{1} d\ell \frac{1}{1+\ell^{2}\xi^{2}}= 2\frac{\tan^{-1}(\xi)}{\xi},
\end{equation}
giving us the result:
\begin{equation}
\langle N_{pairs}\rangle=C_{n}\frac{8\pi c}{|n_{g1}-n_{g2}|L_{z}}\frac{\tan^{-1}(\xi)}{\xi}.
\end{equation}

The full expression is then:
\begin{equation}
\langle N_{pairs}\rangle = \frac{4\pi\hbar N_{p}|\chi_{\text{eff}}^{(2)}|^{2}}{\epsilon_{0}\lambda_{p}\lambda_{1}\lambda_{2}}\frac{n_{g1}n_{g2}n_{gp}}{n_{p}^{2}n_{1}^{2}n_{2}^{2}}w_{p}^{2}w_{1}^{2}w_{2}^{2}\left(\frac{k_{1}k_{2}k_{p}\xi_{1}\xi_{2}\xi_{p}}{L_{z}(k_{1}\xi_{1}+k_{2}\xi_{2}+k_{p}\xi_{p})}\right)^{2} \frac{8\pi c}{|n_{g1}-n_{g2}|L_{z}}\frac{\tan^{-1}(\xi)}{\xi}.
\end{equation}

To better compare our result with that in \cite{bennink2010optimal}, we make his corresponding substitutions and simplifications. Using the expression (implicitly defining $A_{+}B_{+}$ \eqref{aplusbplus}):
\begin{equation}
\frac{\xi}{A_{+}B_{+}}=\frac{k_{p}^{2}\xi_{1}\xi_{2}\xi_{p}}{(k_{1}\xi_{1} + k_{2}\xi_{2} + k_{p}\xi_{p})^{2}},
\end{equation}
we can simplify $\langle N_{pairs}\rangle$ significantly:
\begin{equation}
\langle N_{pairs}\rangle = \frac{32\pi^{2}\hbar c N_{p}L_{z}|\chi_{\text{eff}}^{(2)}|^{2}}{\epsilon_{0}\lambda_{p}\lambda_{1}\lambda_{2}}\frac{n_{g1}n_{g2}n_{gp}}{n_{p}^{2}n_{1}^{2}n_{2}^{2}|n_{g1}-n_{g2}|}\frac{k_{1}k_{2}}{k_{p}L_{z}}\frac{1}{A_{+}B_{+}} \tan^{-1}(\xi),
\end{equation}
which gives us the final pair generation probability (equal to the expectation value $\langle N_{pairs}\rangle$):
\begin{equation}\label{OurValue}
\boxed{\langle N_{pairs}\rangle = \frac{64\pi^{3}\hbar c }{\epsilon_{0}}\frac{n_{g1}n_{g2}n_{gp}}{n_{p}^{3}n_{1}n_{2}|n_{g1}-n_{g2}|}\frac{|\chi_{\text{eff}}^{(2)}|^{2}}{\lambda_{1}^{2}\lambda_{2}^{2}}\frac{\tan^{-1}(\xi)}{A_{+}B_{+}}N_{p}}.
\end{equation}
For convenience, we point out:
\begin{equation}\label{aplusbplus}
A_{+}B_{+}=\frac{(k_{1}\xi_{1} + k_{2}\xi_{2} + k_{p}\xi_{p})(k_{1}\xi_{1}(\xi_{2}-\xi_{p}) + k_{2}\xi_{2}(\xi_{1}-\xi_{p})+k_{p}\xi_{p}(\xi_{1}+\xi_{2}))}{k_{p}^{2}\xi_{1}\xi_{2}\xi_{p}}.
\end{equation}

Comparing our result \eqref{OurValue} to that found in the Bennink Paper (equation 40 in \cite{bennink2010optimal}), we find the following correction factor:
 \begin{equation}
 \frac{\langle N_{pairs}^{(ours)}\rangle}{\langle N_{pairs}^{(Bennink)}\rangle}=\frac{1}{\epsilon}\frac{n_{g1}n_{g2}n_{gp}}{n_{1}^{2}n_{2}^{2}n_{p}^{2}}.
 \end{equation}
 Note that in \cite{bennink2010optimal}, the factor $\epsilon$ is a nondescript efficiency factor accounting for the loss factor when using quasi-phase matching as well as reflection/absorption losses in the medium. In Bulk SPDC with anti-reflection coatings and no quasi-phase matching, $\epsilon$ will be approximately unity (as absorption over these small distances is generally regarded as small). Depending on the values of the indices of refraction, this illustrates that the absolute generation rate may be as low as ten percent of that predicted by the Bennink formula for incides of refraction above 2.2. The only major differences in Bennink's derivation are in expressing the Hamiltonian in terms of the electric field, and in the general form for the electric field operator. Both of these do not affect the primary results of his paper regarding the qualitative dependencies of the down-converted light and the conditions under which it is optimized.

\subsection{Consistency with formulas in our main tutorial}
 To show that equation \eqref{OurValue} reproduces formula 50 in our tutorial \cite{Schneeloch_2019_RateRef} in the limit of small $\xi$ (i.e., collimated beam limit), we will consider the case of type-II collinear SPDC where $\xi_{1}=\xi_{2}=\xi_{p}$ and that $\xi_{j}$ is small.

 In this regime we have that:
 \begin{subequations}
 \begin{align}
 A_{+}B_{+}&\approx 4\\
 \tan^{-1}(\xi)&\approx \xi\\
 \xi&\approx \xi_{p}=\frac{L_{z}}{4 k_{p}\sigma_{p}^{2}}\\
 \sigma_{1}&=\sigma_{2}=\sigma_{p}\sqrt{2}
 \end{align}
 \end{subequations}
Where (for comparison) the formula for the generation rate in \cite{Schneeloch_2019_RateRef} (equation 50 therein) under our assumptions is:
\begin{equation}
R_{SM}^{t2}=\frac{1}{16\pi \epsilon_{0}c^{2}}\frac{n_{g1}n_{g2}}{n_{1}^{2}n_{2}^{2}n_{p}}\frac{(d_{\text{eff}})^{2}\omega_{p}^{2}}{\Delta n_{g}}\frac{P}{\sigma_{p}^{2}}L_{z},
\end{equation}
the formula \eqref{OurValue} derived in the previous section simplifies to:
\begin{equation}
R_{(revised)}^{t2}\rightarrow\frac{1}{16\pi \epsilon_{0}c^{2}}\frac{n_{g1}n_{g2}n_{gp}}{n_{1}n_{2}n_{p}^{4}}\frac{(d_{\text{eff}})^{2}\omega_{p}^{2}}{\Delta n_{g}}\frac{P}{\sigma_{p}^{2}}L_{z},
\end{equation}
which differs by a factor of order unity:
\begin{equation}
\frac{R_{(revised)}^{t2}}{R_{SM}^{t2}}=\frac{n_{1}n_{2}n_{gp}}{n_{p}^{3}}
\end{equation}
This discrepancy is due to the refined treatment in this discussion. In particular, we no longer approximate the pump as having one discrete momentum mode, and instead fully convert the hamiltonian into an integral over frequency modes prior to obtaining the rate of down-converted photon pairs. 

Because this correction factor arises due to factors \emph{prior} to determining the type of phase matching in the nonlinear medium, it can and should be added to all generation rate formulas in \cite{Schneeloch_2019_RateRef} for improved accuracy, though in practice this may amount to only a few percent change. For all the cases demonstrated experimentally in the body of the main paper \cite{Schneeloch_2019_RateRef}, these correction factors bring the theoretical predictions closer to their experimental values as seen in the following table. Note that estimates of the uncertainties of the group index were taken to match the corresponding estimated uncertainties in the body of the paper.

\begin{table}[h!]
\centering
 \begin{tabular}{| M{1.75cm} | M{4.5cm} | M{4.5cm} | M{4.5cm}|}
 \hline
  & Type-0, SM in PPLN & Type-I, MM in BiBO & Type-II, SM, in PPKTP \\
 \hline
     $n_{gp}$ & $2.292\pm 0.001$ & $1.989\pm0.001$ & $1.811\pm0.002$ \\
        \hline
        $\begin{smallmatrix}
        \text{correction}\\ 
        \text{factor}
        \end{smallmatrix}$ & $1.00648\pm 0.002$ & $1.09166\pm 0.002$ & $1.02996\pm 0.004$ \\
        \hline
         $R_{th}^{(paper)}$ & $94.86\pm10.89\times 10^{6}$/s/mW & $53.87\pm10.87 \times 10^{6}$/s/mW & $23.58\pm5.60 \times 10^{6}$/s/mW \\
         $R_{th}^{(revised)}$ & $95.43\pm10.96\times 10^{6}$/s/mW & $58.81\pm11.87 \times 10^{6}$/s/mW & $24.29\pm5.77 \times 10^{6}$/s/mW \\
          $R_{exp}$ & $95.63\pm2.71\times 10^{6}$/s/mW & $64.68 \pm 1.69 \times 10^{6}$/s/mW & $35.5\pm0.8 \times 10^{6}$/s/mW \\
 \hline
 \end{tabular}
 \caption{Table of additional experimental parameters and effect on results}
 \end{table}

\section{Conclusion}
In these comments, we have derived from first principles the pair generation rate formula for Spontaneous Parametric Down-Conversion (SPDC) from a single pump Gaussian beam into a pair of Gaussian signal/idler collection modes, and compared these results to the seminal treatment by Bennink in \cite{bennink2010optimal} for consistency and completeness. Along the way, we have cleared up a source of confusion between quantization volume, and the dimensions of the nonlinear medium. 

Our new formulas for the pair generation rate differ from those in our tutorial \cite{Schneeloch_2019_RateRef} and in Bennink's treatment \cite{bennink2010optimal} by dimensionless constants (based on indices of refraction) that do not affect previous findings on what conditions will optimize the pair collection rate and heralding efficiency, or on how focusing affects the bandwidth of the down-converted light (among other aspects). 

Moreover, our new formulas, represent a small, order-of-unity correction to what we derived in \cite{Schneeloch_2019_RateRef}, which brings our theoretical predictions closer to our experimentally obtained results. With this formalism laid out in more expanded detail, it is our hope that even more sophisticated calculations of down-converted light will become straightforward to a broader community of quantum scientists.

\begin{acknowledgments}
We gratefully acknowledge support from the Air Force Research Laboratory. The views expressed are those of the authors and do not reflect the official guidance or position of the United States Government, the Department of Defense or of the United States Air Force. The appearance of external hyperlinks does not constitute endorsement by the United States Department of Defense (DoD) of the linked websites, or of the information, products, or services contained therein. The DoD does not exercise any editorial, security, or other control over the information you may find at these locations.
\end{acknowledgments}

\bibliography{EPRbib16}

\end{document}